 \documentclass{vgtc}                          

\ifpdf
  \pdfoutput=1\relax                   
  \usepackage{graphicx}                
  \DeclareGraphicsExtensions{.pdf,.png,.jpg,.jpeg} 
\else
  \usepackage{graphicx}                
  \DeclareGraphicsExtensions{.eps}     
\fi%

\graphicspath{{figures/}{pictures/}{images/}{./}} 

\usepackage{microtype}                 
\PassOptionsToPackage{warn}{textcomp}  
\usepackage{textcomp}                  
\usepackage{mathptmx}                  
\usepackage{balance}
\usepackage{times}                     
\usepackage{cite}                      
\usepackage{tabu}                      
\usepackage{booktabs}                  
\usepackage{dirtytalk} 
\usepackage{enumitem}
\newcommand\blfootnote[1]{%
  \begingroup
  \renewcommand\thefootnote{}\footnote{#1}%
  \addtocounter{footnote}{-1}%
  \endgroup
}

\onlineid{0}

\vgtccategory{Research}

\vgtcinsertpkg

\title{Co-Designing Unstructured Text Data Visualization Systems}

\author{Beck Langstone\thanks{e-mail: becklangstone@cmail.carleton.ca}\\ %
         \parbox{1.4in}{\scriptsize \centering Department of Human-Computer Interaction\\ Carleton University}
        \and Fateme Rajabiyazdi\thanks{e-mail: fateme.rajabiyazdi@carleton.ca }\\ %
      \parbox{1.4in}{\scriptsize \centering Department of Systems and Computer Engineering \\ Carleton University}}

\abstract{We present our in-progress work on co-designing a visualization tool for presenting unstructured text. We have conducted a focus group with a variety of professionals who regularly analyze large corpora of unstructured text. Our preliminary insights indicate there is an unmet need to visually explore the dynamics between entities and actors extracted from unstructured text. Additionally, large corpora contain multiple perspectives on the same series of events. There is a need to disentangle these perspectives and visually show the multiple narratives present in the data. In our future work, we will co-design low-fidelity prototypes to create a broad consideration space of possible solutions for visualizing unstructured text.

} 

\CCScatlist{
  \CCScatTwelve{Human-centered computing}{Visu\-al\-iza\-tion}{Visu\-al\-iza\-tion techniques}{Text Data Visualization};
  \CCScatTwelve{Human-centered computing}{Visu\-al\-iza\-tion}{Visualization design and evaluation methods}{}
}

\begin{document}



\maketitle

\section{Introduction}
There has been an explosion in the amount of qualitative data generated every day, and it is expected to continue growing. Unstructured text data is ubiquitous, and can be found in open-ended surveys, comments, social media posts, news articles, and more. With the growth in the amount of qualitative data, there’s also an increasing need to understand the data. Finding trends, identifying risks, actors, influencers, and other patterns will become increasingly difficult as the quantity of data increases. Visualizing this data can be a solution to address these challenges. 

\blfootnote{Poster presented at Graphics Interface Conference 2022\\
16-19 May - Montreal, QC, Canada\\
Copyright held by authors.\\ }

In the literature, there are design guidelines for developing quantitative data visualizations. However, there is a lack of understanding and guidelines on how to design and develop qualitative data visualizations.  Additionally, while there are existing qualitative data visualization tools, they are often tailored for a specific goal or use case~\cite{sultanum2021text, 2018_ConceptVector}. For example, Storifier~\cite{sultanum2021text} is a text visualization tool that is designed to support journalists with close reading tasks, but it is not well suited for short bodies of text. ConceptVector~\cite{2018_ConceptVector} is a text visualization tool that allows users to analyze text using interactive lexicon building, but does not allow users to explore the actors, entities and narratives of the text data. There is a need to create a qualitative data visualization tool with a level of generality and customizability that makes it applicable to a wide variety of use cases. 

The objective of this research is to design and develop a qualitative data visualization dashboard that provides users with ways to gain insights from unstructured text data. Our goal is to maintain a level of generality and customization that makes the tool applicable to a variety of use cases. With our in-progress research, we are defining requirements for an unstructured text data visualization tool. Using focus groups, we are gathering requirements to design a solution.

\section{Background and Related Work}

Our review of related work focuses on existing visualization tools and approaches to text visualization. 
ConceptVector~\cite{2018_ConceptVector} is a text analysis visualization tool that allows users to build their own concepts from a set of seed terms. The unique concept building features of ConceptVector were designed to mitigate errors that arise as a result of natural language polysemy and homonymy in other text analysis systems that use pre-built lexicons to define concepts. Polysemy and hononymy both are traits of words that have multiple senses associated with one lexical form, which creates lexical ambiguity. ConceptVector supports the creations of concepts generally, but the concept of a particular actor or entity is not particularly well supported with this tool. We are tackling this issue by designing our tool with the tasks of exploring and understanding entities and actors in mind. There is not a straightforward way to understand multiple storylines as they are reported by differing perspectives, which is a need we aim to address with our design.

Lu et al.~\cite{2018_TopicDrivers} designed a tool to assist users with large-scale analysis of text data aggregated from different media sources, across more than one corpus. Users can explore, link, and annotate data from multiple sources to investigate drivers of discourse. This tool employs a user guided semantic lexical matching scheme, causality metrics for identifying media drivers, and a causality-driven annotation scheme for exploring data. These causality features allow users to see which events may have caused another using a Granger causality test. While this causality metric is unique and interesting, it makes the assumption that causal effects can only occur in one direction, where X causes Y. However, causality can be bidirectional, where X can cause Y and Y can cause X. The relationship between cause and effect can be understood as part of a narrative, and there may be differing narratives in data sources. We aim to address this challenge by supporting multiple storylines in our visualization. 

Storifier~\cite{sultanum2021text} is a text visualization tool that helps journalists and news analysts with the task of close reading. While automated text analysis is used by journalists, they do not \say{replace the human judgement needed for fine-grained, qualitative forms of analysis}. This tool focuses on supporting the task of reading, allowing users to drill down to the original text from a corpus-level overview. Users are able to quickly and easily transition between levels of detail, starting with a summarized view that provides a top-down awareness of corpus trends and high level patterns. Users are able to navigate and explore their dataset with a variety of compoundable filters. In these views, users are able to tag and annotate their data, allowing them to revisit user-defined segments of interest within documents. This ability to explore a large corpus of text data at the macro and micro scale offers a supportive interface to users to efficiently explore narratives. While this tool is well suited for long sections of unstructured text, it is not ideal for analyzing short unstructured text sourced from tweets or other short posts. Our tool will be designed with the flexibility to tackle both forms of unstructured text. 

\section{Methodology}

Our goal is to design and develop a visualization with a level of generality and customization options to create a flexible tool that is applicable to a variety of use cases. To accomplish this goal, we have conducted an initial focus group to gather requirements. In order to build this visualization tool, we have developed our approach on previous research. We are using Kerzner’s Framework for Creative Visualization-Opportunities Workshops~\cite{kerzner2018framework} as an approach to explore and characterise domain problems and visualization focus groups with our industry partners. This general approach has guided our first focus group session successfully, and we are in the process of analysing and coding the transcript data before beginning our next focus group session. 
\par
Due to the pandemic, our focus group session has been conducted online via Microsoft Teams. Sessions were audio and video recorded for analysis. Following Kezner's model, our focus group session was made of 3 main stages: Opening, core, and closing. In the opening, we introduced ourselves to foster agency and encourage collegiality and trust. We were able to establish interest in our area of research, which was shown by our participants' enthusiasm and willingness to share. In our first session, we explored the following questions.
\begin{itemize} [noitemsep,nolistsep]
    \item What are your current analysis methods?
    \item What are your current analysis challenges?
    \item Are there limitations to the existing tools you use?
\end{itemize}

These questions elicited thoughtful responses from our participants, who each had unique backgrounds and experiences.

After a break, we moved on to a wishful thinking exercise. In this exercises, we asked participants to imagine they had just been given a new dataset to analyze. Next, we asked the following.
\begin{itemize}[noitemsep,nolistsep]
    \item What would you like to know about the data?
    \item What tasks would you like to do with the data?
    \item Is there a visual representation of the data that could be helpful to you?
\end{itemize}

We are also following guidance from Munzner’s nine stage design study methodology  ~\cite{sedlmair2012design}  to guide our research. At this point, we are in the discover phase where we are focusing on problem characterization and abstraction. This stage is particularly critical to our work, as one of our research goals is to create a visualization tool that is applicable to a variety of use cases. This stage will help us ensure that our abstractions are well supported and transferable to other domains. While the refinement of the abstraction begins in this stage, it will continue through all subsequent stages of this model. 

\section{Preliminary Findings}
From our initial focus group session, we have uncovered themes that will inform the design of our visualization tool. Two main tasks our participants are trying to accomplish with their data analysis are uncovering narratives and storytelling. Often, our participants will begin their analysis with a research question that helps inform how they go about their data analysis. Their data analysis is exploratory in nature, which presents an opportunity for supporting exploration with visualization.

Participants expressed that they experience difficulty in analyzing and understanding the relationships between entities in text over time, and from differing perspectives. While they are able to extract entities from unstructured text, a significant challenge is understanding the source's perspective and bias. In large, multi-source data sets of unstructured text there are inevitably multiple perspectives, leading to multiple entangled storylines. The dynamics of these entities are made complex by these two aspects: the changes of these dynamics over time, and the way these dynamics are described by different sources. There is a need to visually explore these dynamics to understand the relationships between entities and actors.

\section{Conclusion and Future Work}
As we continue to characterize the challenges of visualizing text data and refine abstractions, we will conduct further focus group sessions. The design requirements gathered from the first focus group will help us design lo-fi prototypes to validate our data abstractions and test visual encoding and interaction mechanisms. As Munzer points out, \say{While there is usually not a \textit{best} solution, there are many \textit{good} and \textit{ok} ones.}~\cite{sedlmair2012design}. By creating multiple lo-fi prototypes, we will create a broad consideration space of possible solutions.

\acknowledgments{This work was supported in part by NSERC and OCE grants and our industry partner, Novacene.ai. We'd like to thank our study participants for their contributions.}
\balance

\bibliographystyle{abbrv-doi}
\bibliography{template}

\begin{thebibliography}{1}

\bibitem{kerzner2018framework}
E.~Kerzner, S.~Goodwin, J.~Dykes, S.~Jones, and M.~Meyer.
\newblock A framework for creative visualization-opportunities workshops.
\newblock {\em IEEE transactions on visualization and computer graphics},
  25(1):748--758, 2018. doi: {{%
10\hspace{.1pt}\discretionary{.}{%
}{.}\hspace{.4pt}1109\discretionary{/}{%
}{/}TVCG\hspace{.1pt}\discretionary{.}{%
}{.}\hspace{.4pt}2018\hspace{.1pt}\discretionary{.}{%
}{.}\hspace{.4pt}2865241}}


\bibitem{2018_TopicDrivers}
Y.~Lu, H.~Wang, S.~Landis, and R.~Maciejewski.
\newblock A visual analytics framework for identifying topic drivers in media
  events.
\newblock {\em IEEE Transactions on Visualization and Computer Graphics},
  24(9):2501--2515, Sept. 2018. doi: {{%
10\hspace{.1pt}\discretionary{.}{%
}{.}\hspace{.4pt}1109\discretionary{/}{%
}{/}TVCG\hspace{.1pt}\discretionary{.}{%
}{.}\hspace{.4pt}2017\hspace{.1pt}\discretionary{.}{%
}{.}\hspace{.4pt}2752166}}


\bibitem{2018_ConceptVector}
D.~Park, S.~Kim, J.~Lee, J.~Choo, N.~Diakopoulos, and N.~Elmqvist.
\newblock {ConceptVector}: Text visual analytics via interactive lexicon
  building using word embedding.
\newblock {\em IEEE Transactions on Visualization and Computer Graphics},
  24(1):361--370, Jan. 2018. doi: {{%
10\hspace{.1pt}\discretionary{.}{%
}{.}\hspace{.4pt}1109\discretionary{/}{%
}{/}TVCG\hspace{.1pt}\discretionary{.}{%
}{.}\hspace{.4pt}2017\hspace{.1pt}\discretionary{.}{%
}{.}\hspace{.4pt}2744478}}


\bibitem{sedlmair2012design}
M.~Sedlmair, M.~Meyer, and T.~Munzner.
\newblock Design study methodology: Reflections from the trenches and the
  stacks.
\newblock {\em IEEE transactions on visualization and computer graphics},
  18(12):2431--2440, 2012. doi: {{%
10\hspace{.1pt}\discretionary{.}{%
}{.}\hspace{.4pt}1109\discretionary{/}{%
}{/}TVCG\hspace{.1pt}\discretionary{.}{%
}{.}\hspace{.4pt}2012\hspace{.1pt}\discretionary{.}{%
}{.}\hspace{.4pt}213}}


\bibitem{sultanum2021text}
N.~Sultanum, A.~Bezerianos, and F.~Chevalier.
\newblock Text visualization and close reading for journalism with storifier.
\newblock In {\em 2021 IEEE Visualization Conference (VIS)}, pp. 186--190.
  IEEE, 2021. doi: {{%
10\hspace{.1pt}\discretionary{.}{%
}{.}\hspace{.4pt}1109\discretionary{/}{%
}{/}VIS49827\hspace{.1pt}\discretionary{.}{%
}{.}\hspace{.4pt}2021\hspace{.1pt}\discretionary{.}{%
}{.}\hspace{.4pt}9623264}}


\end{thebibliography}

\end{document}